\documentclass[3p,times]{elsarticle}
\usepackage{ecrc}
\volume{00}
\firstpage{1}
\journalname{Nuclear Physics A}
\runauth{G. Vujanovic, J.-F. Paquet, G. S. Denicol, M. Luzum, B. Schenke, S. Jeon, C. Gale}
\jid{npa}
\jnltitlelogo{Nuclear Physics A}
\usepackage{amssymb}
\biboptions{square,comma,numbers,sort&compress}
\usepackage[figuresright]{rotating}

\begin{document}
\begin{frontmatter}
\dochead{}
\title{Probing the early-time dynamics of relativistic heavy-ion collisions with electromagnetic radiation}
\author{Gojko Vujanovic$^1$, Jean-Fran{\c c}ois Paquet$^1$, Gabriel S. Denicol$^1$, Matthew Luzum$^{1,2}$, Bj{\"o}rn Schenke$^3$, Sangyong Jeon$^1$, and Charles Gale$^{1,4}$}
\address{$^1$Department of Physics, McGill University, 3600 University Street, Montreal, QC H3A 2T8, Canada\\ $^2$Lawrence Berkeley National Laboratory, Berkeley, California 94720, USA\\ $^3$Physics Department, Bldg. 510A, Brookhaven National Laboratory, Upton, NY 11973, USA\\ $^4$Frankfurt Institute for Advanced Studies, Ruth-Moufang-Str. 1, D-60438 Frankfurt am Main, Germany}
\begin{abstract}
Using 3+1D viscous relativistic fluid dynamics, we show that electromagnetic probes are sensitive to the initial conditions and to the out-of-equilibrium features of relativistic heavy-ion collisions.  Within the same approach, we find that hadronic observables show a much lesser sensitivity to these aspects. We conclude that electromagnetic observables allow access to dynamical regions that are beyond the reach of soft hadronic probes.
\end{abstract}
%\begin{keyword}
%\end{keyword}
\end{frontmatter}

\section{Introduction}

Modern relativistic heavy ion colliders, such as the Relativistic Heavy Ion Collider (RHIC) and the Large Hadron Collider (LHC), are capable of exciting strongly interacting matter to the point where its partonic degrees of freedom become available for experimental investigations. These degrees of freedom form a new state of matter: the Quark Gluon Plasma (QGP). Currently, relativistic viscous hydrodynamics provides a good description of the time evolution of the QGP and of the hadronic medium. Owing to their penetrating nature, electromagnetic (EM) probes are sensitive to the {\it entire} time evolution of the medium created by heavy ion collisions, which  allows for a more stringent verification of the hydrodynamical behavior and also imposes more severe constraints on the parameters of the simulation approaches. In this study the sensitivity of EM observables to the initial conditions and to the shear relaxation time of viscous hydrodynamics is investigated. 

\section{Viscous hydrodynamics}
We concentrate on conditions germane to RHIC. 
To describe the medium produced in nuclear collisions at such energies, we use \textsc{music}, a 3+1D hydrodynamical simulation \cite{Schenke:2010rr}. The initial configurations are set by  the optical Glauber model tuned to hadronic observables. The evolution of the energy-momentum tensor, $T^{\mu \nu }$, is governed by the conservation equation, $\partial _{\mu }T^{\mu \nu }=0$. Note that $T^{\mu \nu }=\varepsilon\, u^{\mu }u^{\nu }-\Delta ^{\mu \nu }P+\pi^{\mu \nu }$, with $\varepsilon $ being the energy density, $P$ the thermodynamic pressure, $u^{\mu }$ the fluid four-velocity, $\pi ^{\mu \nu }$ the shear-stress tensor, and $\Delta ^{\mu \nu }=g^{\mu \nu }-u^{\mu }u^{\nu}$ the projection operator onto the 3-space orthogonal to the velocity. 

The dynamics of the shear-stress tensor are given by a version of Israel-Stewart (I-S) theory \cite{Israel1976310,Israel:1979wp}:
\begin{equation}
\Delta _{\alpha \beta }^{\mu \nu }u^{\lambda}\partial_{\lambda}\pi ^{\alpha\beta }+\frac{4}{3}\pi^{\mu \nu }\partial _{\lambda }u^{\lambda }= \left( \pi^{\mu\nu}_{\rm NS} - \pi^{\mu\nu} \right) / \tau_\pi
\end{equation}
where the Navier-Stokes limit of the shear-stress tensor is $\pi^{\mu \nu}_{\rm NS} = 2 \eta \,\sigma ^{\mu \nu }= 2 \eta \Delta _{\alpha \beta }^{\mu \nu }\partial^{\alpha }u^{\beta }$, with $\Delta _{\alpha \beta }^{\mu\nu }=\left( \Delta _{\alpha }^{\mu }\Delta _{\beta }^{\nu }+\Delta _{\beta}^{\mu }\Delta _{\alpha }^{\nu }\right) /2-\Delta _{\alpha \beta }\Delta^{\mu \nu }/3$ being the double, symmetric, traceless projection operator. There are two coefficients, the shear viscosity $\eta $, also present in Navier-Stokes theory, and the shear relaxation time, $\tau _{\pi }$, which only exists in Israel-Stewart theory. These are the only terms considered in this study.  Furthermore, we assume the existence of an effective shear viscosity coefficient that is proportional to the entropy density: $\eta / s = 1/4 \pi$. The relaxation time is assumed to be of the form $\tau_{\pi}=b_{\pi}\left[\frac{\eta}{\varepsilon +P}\right]$, and we will choose here $b_{\pi }=3,5,$ and $10$. Physically, $\tau_\pi$ governs the rate at which $\pi^{\mu\nu}$ evolves  and relaxes towards the Navier-Stokes value. The coefficient $b_\pi$ is constrained by causality  to $b_{\pi }\geq4/\left[3\left( 1-c_{s}^{2}\right)\right]$, where $c_{s}$ is the velocity of sound \cite{Denicol:2008ha}. 

To investigate the sensitivity of EM probes to the initial conditions of the medium, we start the fluid dynamic evolution in and out of equilibrium by introducing an initial $\pi^{\mu\nu}_0=c\times2\eta\sigma^{\mu\nu}$ where $c=0,2$, and $\sigma^{\mu\nu}$ is computed using initial flow $u^\mu_0=\left(\cosh \eta_s,0,0,\sinh \eta_s \right)$, with the space-time rapidity given by $\eta_s=(1/2)\ln\left[(t+z)/(t-z)\right]$ where $t$ is time and $z$ is the longitudinal coordinate. A practical set of coordinates is hyperbolic space-time variables: $\tau=\sqrt{t^2-z^2}$ and $\eta_s$; in these coordinates $u^\mu_0 = (1, 0, 0, 0)$. 

\section{Electromagnetic production rates and their viscous corrections}
Viscous corrections to EM thermal production rates are introduced by including asymmetric corrections of the form $\delta n=C\,n(p)(1\pm n(p))p^\alpha p^\beta \pi_{\alpha\beta}/[2T^2(\varepsilon+P)]$ to the spherically-symmetric, thermal distribution functions $n (p) $ present in the rate calculations \cite{dion-paquet-schenke-young-jeon-gale}. The constant $C$ may be species-dependent \cite{Dusling:2009df}; here we shall set $C = 1$. With this formulation, thermal rates become dependent on the out-of-equilibrium hydro-evolution of $T^{\mu\nu}$.  For dileptons, we use the quark-antiquark annihilation rate into dileptons at leading order (Born approximation) to describe the virtual photon emission of the QGP phase. In the hadronic sector, our rates are based on the Eletsky {\it et al.} \cite{eletsky-belkacem-ellis-kapusta, martell-ellis,vujanovic-ruppert-gale} forward scattering amplitude model, where vector mesons interact with a bath of pions and nucleons. The Vector Dominance Model (VDM) \cite{gounaris-sakurai} couples the vector mesons to virtual photons. Viscous corrections to these rates are presented in \cite{Vujanovic:2013jpa}. For photons, the emission from the QGP sector consists of 2-to-2 processes $q+\bar q\rightarrow g+\gamma$ and $q (\bar q)+g\rightarrow q (\bar q) +\gamma$, along with their Hard Thermal Loop corrections \cite{Kapusta:1991qp}. The QGP photon rates have been extended to include viscous anisotropic momentum distributions. The hadronic medium (HM) photon-producing reactions are described by kinetic theory, using Massive-Yang-Mills Lagrangian meson-meson interactions \cite{Turbide:2003si}. Viscous extensions are derived in \cite{dion-paquet-schenke-young-jeon-gale}. 

\section{Results}

\begin{figure}[!th]
\begin{center}
\begin{tabular} {c c}
\includegraphics[width=0.45\textwidth]{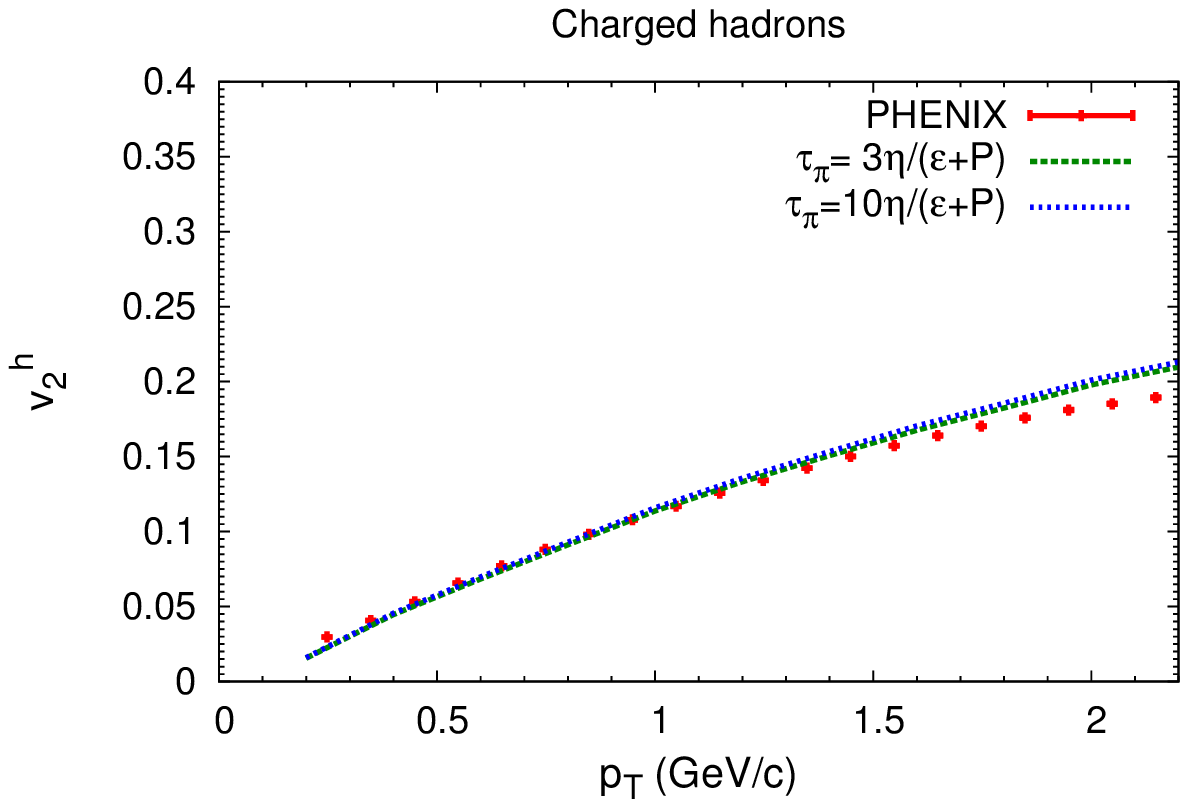} \includegraphics[width=0.45\textwidth]{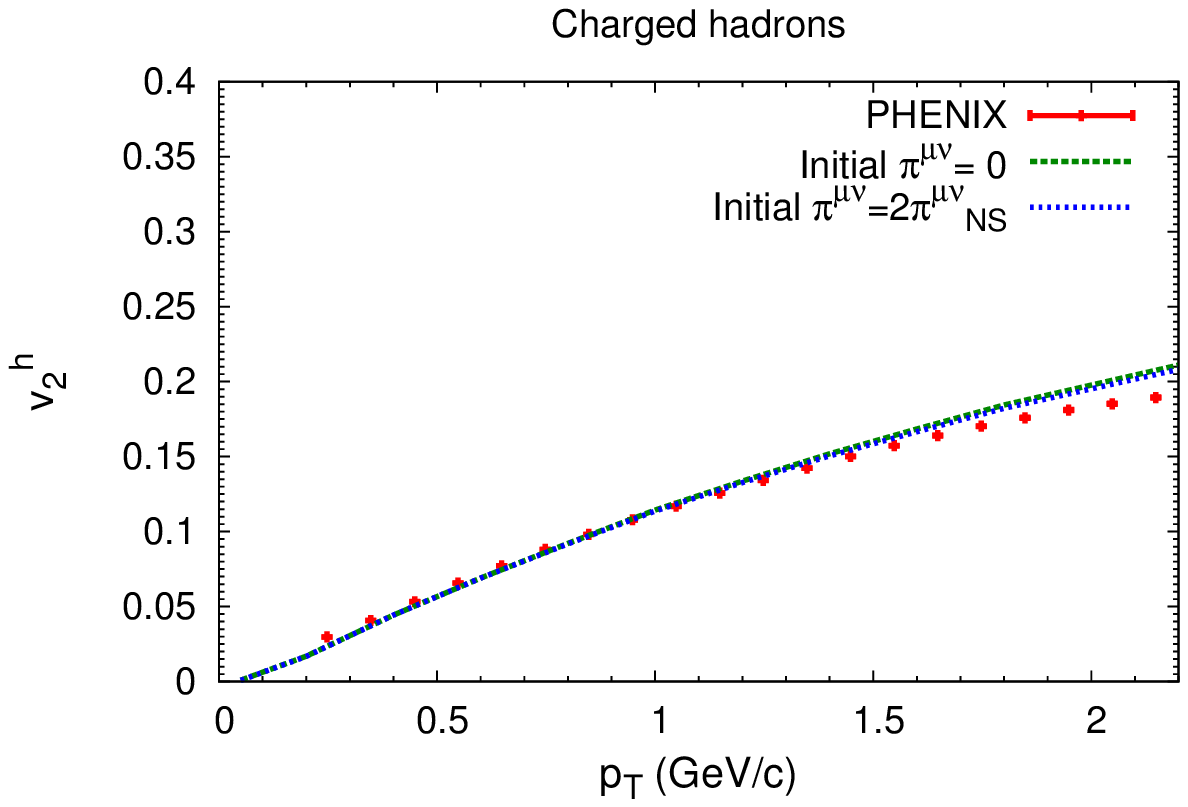} 
\end{tabular}
\end{center}
\caption{The effects of $\tau_\pi$ (left panel) and $\pi^{\mu\nu}_0$ (right panel) on elliptic flow of charged hadrons created in collisions of Au + Au at 200 $A$ GeV, in the  20--40\% centrality class. In the left panel we set $\pi^{\mu\nu}_0 = 0$, and in the right panel, $\tau_\pi = 5 \eta/(\varepsilon + P)$. }
\label{fig:v2_ch}
\end{figure}

It is  important to first verify whether the new hydrodynamic parameter space explored in this work modifies the interpretation of hadronic measurements performed over several years.  Charged hadrons are expected to have a  limited sensitivity to the  early dynamics of the strongly interacting medium created in heavy ion collisions because they are created at the hydrodynamic freeze-out hyper-surface. Our calculations do show that the variations in $\tau_\pi$ and $\pi^{\mu\nu}_0$ studied here have little effect on the $v_2$ of charged hadrons, and this is illustrated in Fig. \ref{fig:v2_ch}.  

\begin{figure}[!th]
\begin{center}
\begin{tabular} {c c c}
\includegraphics[width=0.45\textwidth]{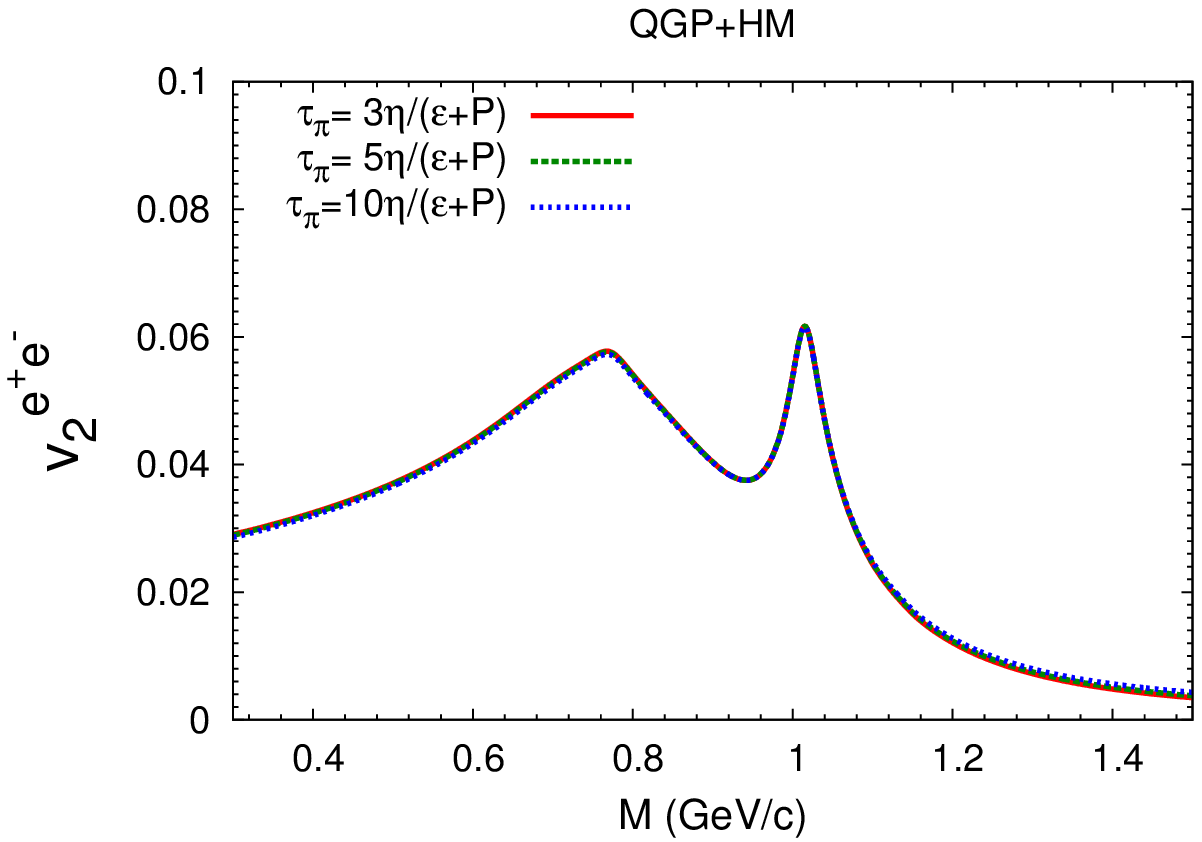}
\includegraphics[width=0.45\textwidth]{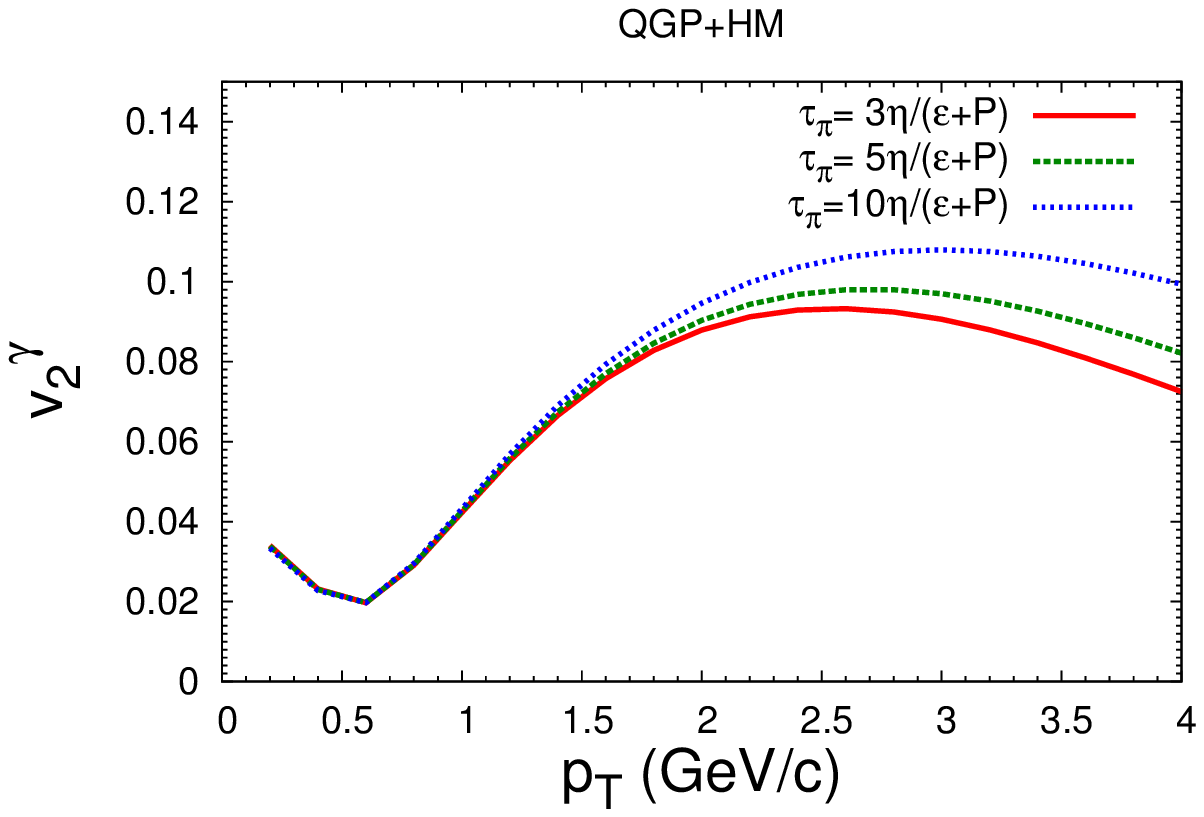}
\end{tabular}
\end{center}
\caption{The effects of $\tau_\pi$ on EM probes for collisions of Au + Au at 200 A GeV, in the 20Ð40\% centrality class. The dilepton (left panel) and photon (right panel) net elliptic flows are shown, for a choice of $\pi^{\mu\nu}_0 = 0$ and different values of the shear relaxation time $\tau_{\pi}$. }
\label{fig:tau_pi}
\end{figure}

%\subsection{Effects of $\tau_\pi$ and $\pi^{\mu\nu}_0$ on elliptic flow of electromagnetic probes}

We find that the elliptic flow of photons is indeed considerably more sensitive to the relaxation time of viscous hydrodynamics than that of hadrons. This sensitivity is significantly reduced for dileptons. Recall that $\tau_\pi$ is in effect the relaxation time of the shear pressure tensor to its Navier-Stokes value. Fixing $\eta/s$ and starting from $\pi^{\mu\nu}_0=0$, a large $\tau_\pi$ will therefore postpone the development of viscous hydrodynamics. In this limit, the value of the elliptic flow of EM probes should become closer to what it is for inviscid hydrodynamics, i.e. $v_2$ should be larger. This is was is seen in these estimates, and this is shown in Figure \ref{fig:tau_pi}, where the net elliptic flow (i.e. that coming from both QGP and HM phases) is shown. Interestingly, the effect is more pronounced for thermal photons than it is for thermal dileptons. Thermal dilepton radiation is dominated by HM in the low invariant mass sector, hence the larger effects of $\tau_\pi$ on the QGP dilepton $v_2$ \cite{Vujanovic:prep} are barely visible in the net thermal invariant mass spectrum. For thermal photons, the turnover in $v_2$ for higher values of $p_T$ shown in Fig. \ref{fig:tau_pi} is QGP-driven\footnote{Note however that pQCD photons are not included in this study and they will have a non-negligible effect at transverse momenta of $\sim$ 3 GeV.}. As mentioned previously, the initial coordinate-space configuration is chosen from an optical Glauber calculation: an upcoming study \cite{Vujanovic:prep} will re-examine the effects studied here with event-by-event fluid dynamics and with IP-Glasma initial states \cite{Gale:2012in}. With these cautionary words in mind, the right panel of Figure \ref{fig:tau_pi} shows that varying the shear relaxation time within the limits considered here leads to an increase of thermal photon elliptic flow of ~30\% at $p_T = 3$ GeV.

%\subsection{Effects of initial $\pi^{\mu\nu}$ on elliptic flow of electromagnetic probes}
\begin{figure}[!t]
\begin{center}
\begin{tabular} {c c c}
\includegraphics[width=0.45\textwidth]{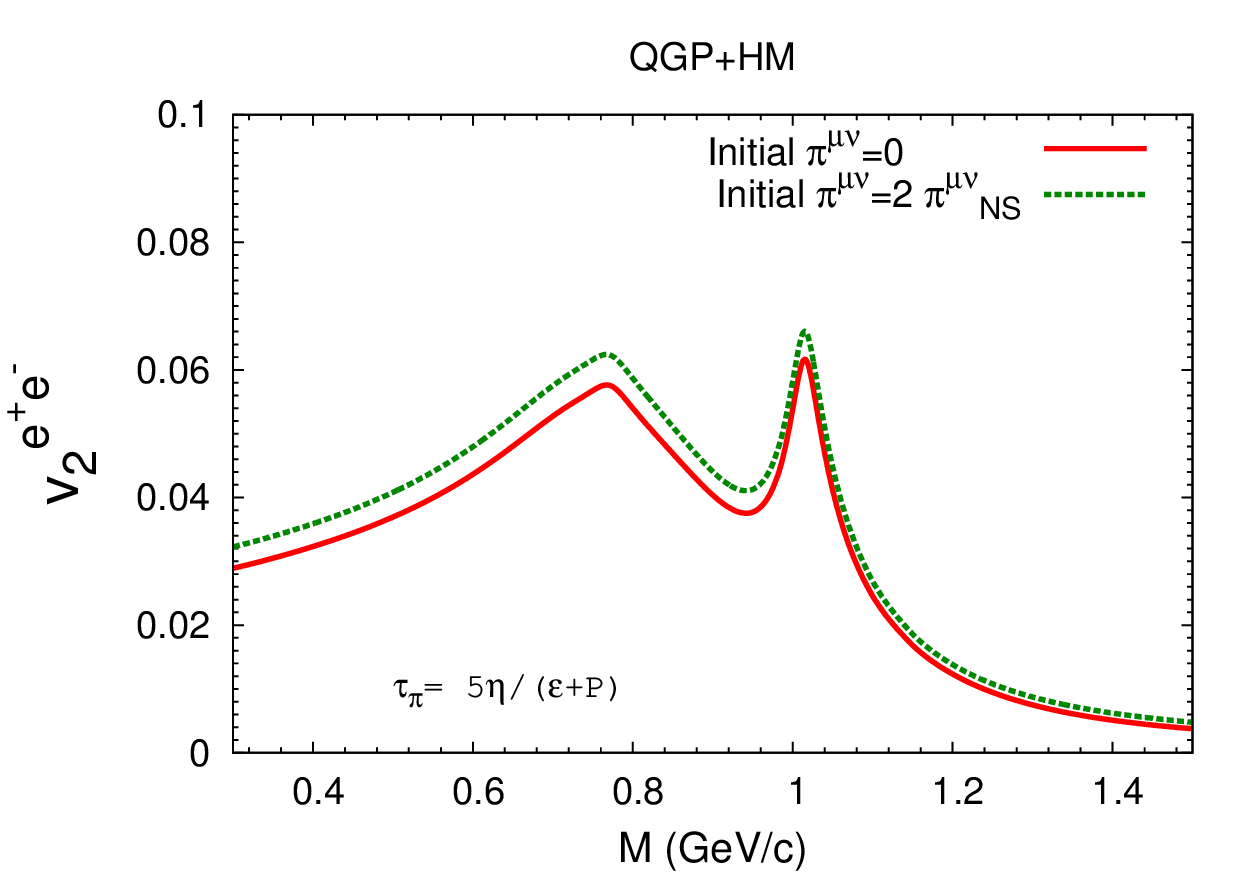} 
\includegraphics[width=0.45\textwidth]{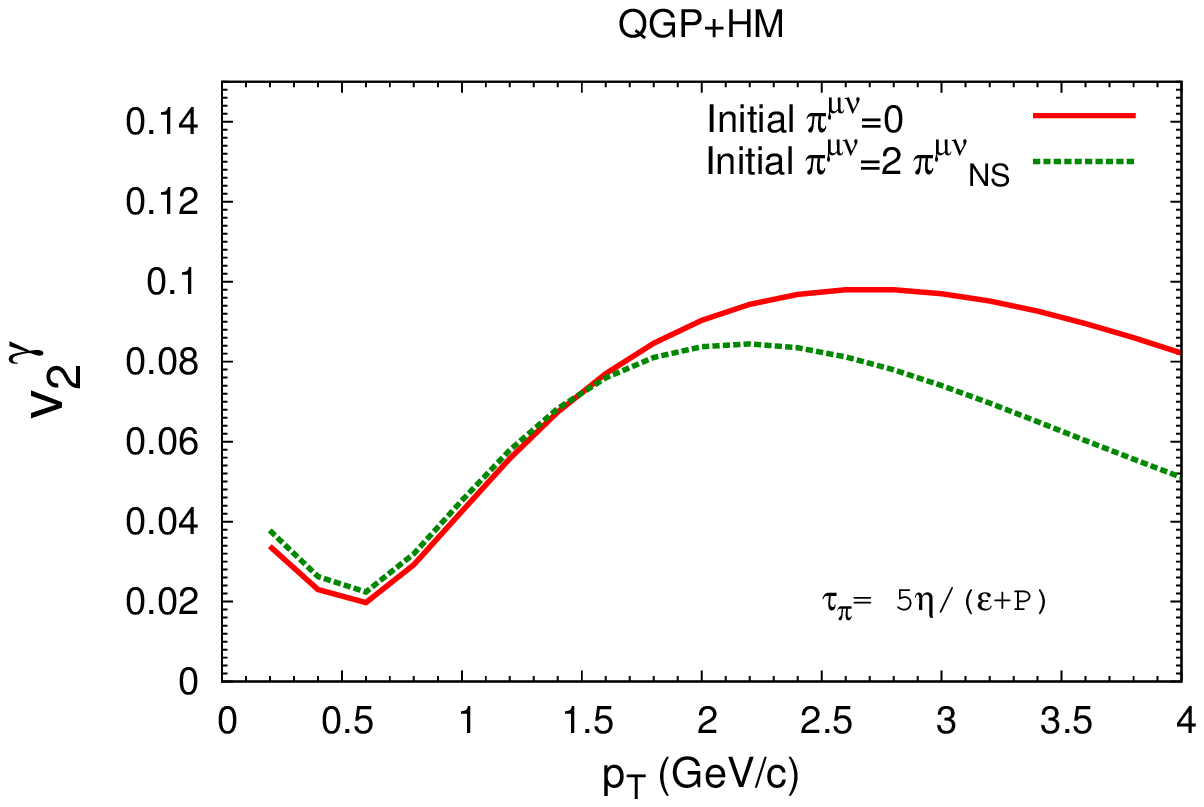}
\end{tabular}
\end{center}
\caption{The effects of $\pi^{\mu\nu}_0$ on EM probes for 20--40\% centrality class at RHIC. The left panel shows the dilepton elliptic flow whereas the right panel shows that of real photons.}
\label{fig:init_pi}
\end{figure}

In what concerns variations of the initial viscous shear pressure tensor, we find that the elliptic flow of dileptons increases regardless of the phase of origin. Figure \ref{fig:init_pi} shows the net elliptic flow of dileptons; this includes lepton pairs emitted from both phases. The hydro evolution does provide the dynamics that supports this interpretation: during the first few fm/c -- when the elements of the shear pressure tensor go through their maxima  -- viscous hydrodynamics introduces a non-linear coupling between the large longitudinal gradients and the transverse gradients by reducing the longitudinal pressure and augmenting the transverse one \cite{Romatschke:2007mq}. Increasing $\pi^{\mu\nu}_0$ (especially $\pi^{zz}_0$) therefore enhances the pressure transfer rate. The $v_2$ of EM probes from the QGP is thus expected to increase because of a non-vanishing value of the initial shear-stress tensor. The effect  is observed in the case of dileptons but the reverse is observed in the case of photons. This can be understood as follows: the integration over momenta needed to get to a dilepton invariant mass distribution suppresses the effect of the asymmetric viscous correction, $\delta n (p)$: the dilepton invariant mass spectra thus only carry a signature of viscosity through the bulk evolution. The photon transverse momentum yield, on the other hand, will be proportional to $p^\mu p^\nu \pi_{\mu \nu}$. The photons therefore also feel the consequence of the $\delta n (p)$ viscous correction which is known to reduce the elliptic flow \cite{Dusling:2009bc,dion-paquet-schenke-young-jeon-gale,Chaudhuri:2011up,Shen:2014cga}. Those effects partially cancel in the case of real photons. Therefore, dilepton (invariant mass) and photon elliptic flow carry complementary information in what concerns the out-of-equilibrium physics in play. This illustrates well the richness of the fluid dynamical problem; a careful analysis is needed to extract all of the physics.

\section{Conclusion}
Our calculations show that EM probes are indeed sensitive to the initial conditions of hydrodynamics, i.e. $\pi^{\mu\nu}_0$ in this work, as well as to the early-time dynamics as represented by the value of the shear relaxation time, $\tau_\pi$. Within the parameter space adopted in this study, the charged hadrons are shown to be mostly unaffected by these aspects, as their characteristics are determined by the conditions existing at the hydrodynamic freeze-out hyper-surface. A more detailed account of our explorations will include the virtual photon emissivities beyond the Born rate, and will feature IP-Glasma initial states along with  other aspects not considered here. Importantly, this work reasserts  the vast potential of electromagnetic radiation as a penetrating probe of hot and dense strongly interacting matter. In more practical terms, photons and dileptons open up a window in the fluid dynamical evolution that has remained closed for hadrons. This in turn may well entail a recalibration of the parameters that currently constitute the hydrodynamics modelling paradigm of relativistic heavy-ion collisions. 

\bibliographystyle{elsarticle-num}
\bibliography{references}

\end{document}